\documentclass[aps,nofootinbib,floatfix,amsmath,amssymb,preprint]{revtex4-1}
\usepackage{preamble}

\newcommand{\editor}[1]{\textnormal{[#1]}}

\begin{document}

\title{The little-hierarchy problem is a little problem: understanding the difference between the big- and little-hierarchy problems with Bayesian probability}
\author{Andrew Fowlie}
\email{Andrew.Fowlie@KBFI.ee}
\address{National Institute of Chemical Physics and Biophysics, Ravala 10, Tallinn 10143, Estonia}
\date{\today}

\begin{abstract}
Experiments are once again under way at the LHC.
This time around, however, the mood in the high-energy physics community is pessimistic.
There is a growing suspicion that naturalness arguments that predict new physics near the weak scale are faulty and that prospects for a new discovery are limited.
We argue that such doubts originate from a misunderstanding of the foundations of naturalness arguments.
In spite of the first run at the LHC, which aggravated the little-hierarchy problem, there is no cause for doubting naturalness or natural theories.
Naturalness is grounded in Bayesian probability logic --- it is not a scientific theory and it makes no sense to claim that it could be falsified or that it is under pressure from experimental data.
We should remain optimistic about discovery prospects; natural theories, such as supersymmetry, generally predict new physics close to the weak scale.
Furthermore, from a Bayesian perspective, we briefly discuss 't Hooft's technical naturalness and a contentious claim that the little-hierarchy problem hints that the Standard Model is a fundamental theory.
\end{abstract}
\maketitle

\section{Introduction}
This summer, after a two-year hiatus, collisions resumed at the LHC at a centre-of-mass energy of \roots{13}. Unlike in the first run, the mood in the high-energy physics community is gloomy. The optimism that characterized the first run \see{Nath:2010zj,Kane:2008zz} was dampened by evidence that there is no new physics near the weak scale and the suspicion that faith in the principle of naturalness was misplaced\cite{Pich:2015tqa,Strumia:2015dca}. This gloom is not universal; there is a rift in the community and a few remain upbeat\cite{Ellis:2014cga,Baer:2015fsa}.

The principle of naturalness emerged in high-energy physics in the late 1970s when Weinberg\cite{Weinberg:1975gm,Weinberg:1979bn}, Susskind\cite{Susskind:1978ms} and Gildener\cite{Gildener:1976ai}, amongst others, identified a ``naturalness'' problem concerning the mass of a fundamental scalar field. The natural scale for such a mass in an effective theory is that at which microscopic physics is important, the cut-off scale, because of quadratic quantum corrections. In the Standard Model (SM), the mass of a complex scalar field determines the weak scale\cite{Glashow:1961tr,Salam:1968rm,Weinberg:1967tq}. Thus, the cut-off scale ought to be close to the weak scale, or else the SM must be fine-tuned. However, without new physics, we expect the cut-off scale in the SM to be around the Planck scale. This dichotomy became known as the ``hierarchy problem'' in reference to the hierarchy between the weak scale and the Planck scale.

As an example, consider the physical mass of a scalar field in an effective scalar-field theory. The physical mass is the pole in the two-point function; diagrammatically,
\begin{align*}
\label{eq:mass}
\begin{tikzpicture}[baseline={([yshift=-0.6ex]current bounding box.center)}]
\node {
\begin{fmffile}{etwopoint}
\begin{fmfgraph*}(40,40)
\fmfleft{i1}
\fmfright{i2}
\fmf{vanilla}{i1,v1}
\fmf{vanilla}{i2,v1}
\fmfv{d.sh=circle,d.f=shaded,d.si=.5w,l.d=0}{v1}
\end{fmfgraph*}
\end{fmffile}
};
\end{tikzpicture}
& = %
\begin{tikzpicture}[baseline={([yshift=-0.6ex]current bounding box.center)}]
\node {
\begin{fmffile}{eline}
\begin{fmfgraph*}(40,40)
\fmfleft{i1}
\fmfright{i2}
\fmf{vanilla}{i1,i2}
\end{fmfgraph*}
\end{fmffile}
};
\end{tikzpicture}
+ %
\begin{tikzpicture}[baseline={([yshift=-0.6ex]current bounding box.center)}]
\node {
\begin{fmffile}{eloop}
\begin{fmfgraph*}(40,40)
\fmfleft{i1}
\fmfright{i2}
\fmf{vanilla}{i1,v1,i2}
\fmf{vanilla}{v1,v1}
\end{fmfgraph*}
\end{fmffile}
};
\end{tikzpicture}
+ \cdots
\end{align*}
The loop results in a quadratic correction to the physical mass:
\begin{align}
m_{\text{Phys}}^2 &= m_0^2(\Lambda) + \frac{\lambda_0}{32\pi^2} \Lambda^2 
+ \cdots
\end{align}
If the physical mass is to be much smaller than the cut-off, $m_{\text{Phys}}^2 \ll \Lambda^2$, we require miraculous cancellations between the bare mass and the quadratic correction. That this is a problem hinges upon a notion of ``naturalness'' --- a theory is natural if its generic prediction for the weak scale is correct, and fine-tuned otherwise.

In the ensuing decades, the hierarchy problem shaped thinking in theoretical physics and precipitated phenomenological interest in so-called natural theories. Popular natural theories included supersymmetric theories\cite{Salam:1974yz,Haber:1984rc,Nilles:1983ge}, in which quadratic corrections to scalar masses vanish\cite{Witten:1981nf}; technicolor theories\cite{Jackiw:1973tr,Cornwall:1973ts,Eichten:1974et,Weinberg:1975gm,Susskind:1978ms,Weinberg:1979bn}, in which there are no fundamental scalars at the weak scale; and large extra dimensions\cite{ArkaniHamed:1998nn,ArkaniHamed:1998rs}, in which the Planck scale is close to the weak scale.\footnote{Kaplan \etal recently presented a novel class of natural theories --- relaxion theories\cite{Graham:2015cka} --- in which a back-reaction to electroweak symmetry breaking enforces a small weak scale.}

However, in the years following LEP experiments \see{Assmann:2002th,Barate:2003sz}, beginning around the year 2000, a ``little-hierarchy problem'' emerged \see{Barbieri:2000gf}. Since LEP forbade new physics close to the weak scale, there must be a little hierarchy between the weak scale and the scale of new physics, and the weak scale might require ``fine-tuning'' such that it is less than  the scale of new physics. This problem was exacerbated in the last few years by the first run of LHC experiments \see{Bechtle:2015nta,Strumia:2011dv}. The discovery of a Higgs boson in 2012\cite{Aad:2012tfa,Chatrchyan:2012ufa} suggested that there is a fundamental scalar near the weak scale, but the absence of new physics suggested that the cut-off of the SM is not near the weak scale. As a result, there is increasing suspicion that naturalness was a flawed criteria \see{Dine:2015xga}.

Indeed, there are at least two common responses to the little-hierarchy problem:
\begin{enumerate}
\item The SM is not an effective theory; quadratic divergences are an unphysical artefact of regularization. With modifications to explain gravity, dark matter and remedy Landau poles, it describes arbitrarily microscopic scales without enormous radiative corrections to the weak scale.
\item The SM is not a natural theory. This is not a problem; naturalness is an aesthetic principle --- an unreliable, prejudiced criteria on which to build knowledge, which was falsified by collider experiments, as demonstrated by the little-hierarchy problem.
\end{enumerate}
These responses are often combined\cite{Salvio:2014soa,Giudice:2014tma,Bardeen:1995kv,
Heikinheimo:2013fta,Kannike:2014mia,Gabrielli:2013hma,
1308.0295,1307.8428,1307.5298,1307.3764,1305.0884,1304.4700,1304.4001,
1210.2848,0707.0633,0709.2750,0809.3406}. It is the second response that I wish to dispel. By following Jaynes' Bayesian ``logic of science''\cite{Jaynes:2003} --- and casting the hierarchy problem in Bayesian language --- it is apparent that the little-hierarchy problem is a little problem, and not a cause for doubting naturalness or natural theories. In passing, I will remark, somewhat unfavourably, upon the first response from a Bayesian perspective.

\section{Bayesian probability logic}

Bayesian probability extends absolute truth and falsehood by permitting a numerical measure of degree of belief. We assign numerical measures of our degrees of belief --- ``probabilities'' --- to scientific theories. Although we cannot verify a theory with data, by availing ourselves of Bayes' theorem, we may calculate that a theory is more probable than a rival theory in light of data.

Probability logic is by now textbook material; we follow Jaynes\cite{Jaynes:2003} and Gregory\cite{Gregory}. The aim is to construct a unique system of probability logic that follows from modest desiderata:
\begin{enumerate}
\item\label{D1} Our degree of belief in a proposition can be represented by a single real number.
\item\label{D2} Although our probability logic shall be quantitative, it must qualitatively agree with common sense.
\item\label{D3} Our probability logic shall be consistent in that we require that
  \begin{enumerate}
  \item Every possible approach to a calculation must lead to an identical result.
  \item All relevant evidence must be considered. We cannot, by fiat, omit relevant information.
  \item Every equivalent state of knowledge must lead to an identical result.
  \end{enumerate}
\end{enumerate}
These desiderata entail a unique system of logical operations including products and sums. The proofs are tedious; they consist of compiling exhaustive lists of possible forms for an operation, then rejecting possible forms one by one with the desiderata. Bayes' theorem,
\beq
\pg{A}{B} = \frac{\pg{B}{A} \times \p{A}}{\p{B}},
\eeq
is a result of probability logic. The significance of Bayes' theorem is that it justifies weak inductive syllogisms such as:
\begin{itemize}
\item If $A$ implies $B$ and $B$, then $A$ more probable, and
\item If $A$ implies $B$ and not $A$, then $B$ less probable,
\end{itemize}
as well as strong deductive syllogisms.

We must admit, however, that the character of our probability --- it is assigned to any proposition or scientific theory --- is not in accord with the conventional frequentist picture of probability or the conventional methodology of science. Indeed, Popper, amongst others, offered brief scathing criticisms of Bayesian probability in science\cite{Popper}:
\begin{iquote}
I do not believe that it is possible to construct a concept of the
probability of hypotheses \ldots nothing is gained by replacing the
word ``true'' by the word ``probable'', and the word ``false'' by the word
``improbable''.
\end{iquote}
This scorn was unwarranted. Bayesian probability rehabilitates inductive reasoning, arguably solving Hume's problem of induction\cite{sep-induction-problem}, and Earman\cite{Earman} notes that Bayesianism is now
\begin{iquote}
\ldots the name stitched to the Jolly Roger of a leading school of statistics and what is arguably the leading view amongst philosophers of science concerning the confirmation of scientific hypotheses and scientific inference in general.
\end{iquote}
Without inductive reasoning, a theory is either falsified and thus abandoned or unfalsified and thus viable but nothing more \see{Lakatos}. For example, the SM Higgs boson was unfalsified prior to its discovery and unfalsified after its discovery --- without induction, the discovery at the LHC cannot alter our belief in the SM Higgs boson.

We will apply Bayes' theorem in the form
\beq\label{Eq:Bayes}
\overbrace{\pg{\model}{\data}}^{\text{Posterior}} = \frac{\overbrace{\pg{\data}{\model}}^{\text{Evidence}}}{\underbrace{\p{\data}}_{\text{Normalization}}} \times \overbrace{\p{\model}}^{\text{Prior}}.
\eeq
The posterior is our degree of belief in a theory, \model, in light of the experimental data, \data{} --- we judge a theory with our posterior belief. The posterior is the product of the evidence --- the probability of obtaining our data assuming the theory --- and our prior belief in the theory.


The prior is a controversial yet critical ingredient. If we supply our belief in a theory prior to an experiment, our belief after the experiment is dictated by Bayes' theorem. Bayes' theorem cannot, however, dictate our prior beliefs.  In keeping with our desiderata that identical states of knowledge ought to lead to identical results, priors ought to reflect our state of knowledge or ignorance rather than our subjective belief. In other words, I advocate objective priors. Although we might, in a qualitative manner, restrict permissible priors, there might not be a unique prior that qualitatively reflects our state of knowledge. If our prior beliefs are personal opinions, we cannot deliver objective conclusions, and Bayes' theorem may be inappropriate for science. After all, we want science to deliver objective truths about the world. However, we could reach intersubjective agreement in light of data despite variation in prior beliefs. Indeed, Earman\cite{Earman} argues that:
\begin{iquote}
It is a fact of life that scientists start with different opinions. To try to quash this fact is to miss the essence of scientific objectivity: the emergence of an evidence driven consensus from widely differing initial conditions.
\end{iquote}
Indeed, in many cases the correspondence between a prior and our knowledge may be fuzzy and we must take solace in the fact that priors are ``washed out.'' This is arguably a reflection of scientific practice.

\section{Bayesian model comparison}

The Bayesian approach is that to compare two theories, we ought to calculate the ratio of their probabilities conditioned on all relevant experimental data,
\begin{equation}
\underbrace{\frac{\pg{\model_a}{\data}}{\pg{\model_b}{\data}}}_{\text{Posterior odds}} = \underbrace{\frac{\pg{\data}{\model_a}}{\pg{\data}{\model_b}}}_{\text{Bayes factor}} \times \underbrace{\frac{\p{\model_a}}{\p{\model_b}}}_{\text{Prior odds}}.
\end{equation}
This trivializes model comparison --- we simply favour the model that is most probable.

The fact that we consider a ratio is important: it eliminates an unknown normalization constant in \refeq{Eq:Bayes}. If we were to consider a single theory, the posterior and prior probabilities must equal unity, that is, certainty.  We might instead consider the Bayesian evidence for a single theory, but this is not the quantity of interest --- it is the probability of obtaining the observed data. This point is stressed by Jaynes\cite{Jaynes:2003}:
\begin{iquote}
\ldots it is meaningless to ask how much those facts \editor{the data} tend ``in
themselves'' to confirm or refute $H_0$ \editor{a hypothesis}. Not only the mathematics, but also our innate common
sense (if we think about it for a moment) tell us that we have not asked any definite,
well-posed question until we specify the possible alternatives to $H_0$ \ldots mere improbability, however great, cannot in itself be
the reason for doubting $H_0$.
\end{iquote}
Furthermore, if the evidence is a probability density as function of the data, it is a dimensionful number that depends non-trivially on our choice of parametrization of the data. It makes no sense to ask whether such an evidence is large or small.


\section{Casting the big-hierarchy problem in Bayesian language}\label{Sec:BayesNat}

\begin{figure}[ht]
\centering
\includegraphics[width=0.75\linewidth]{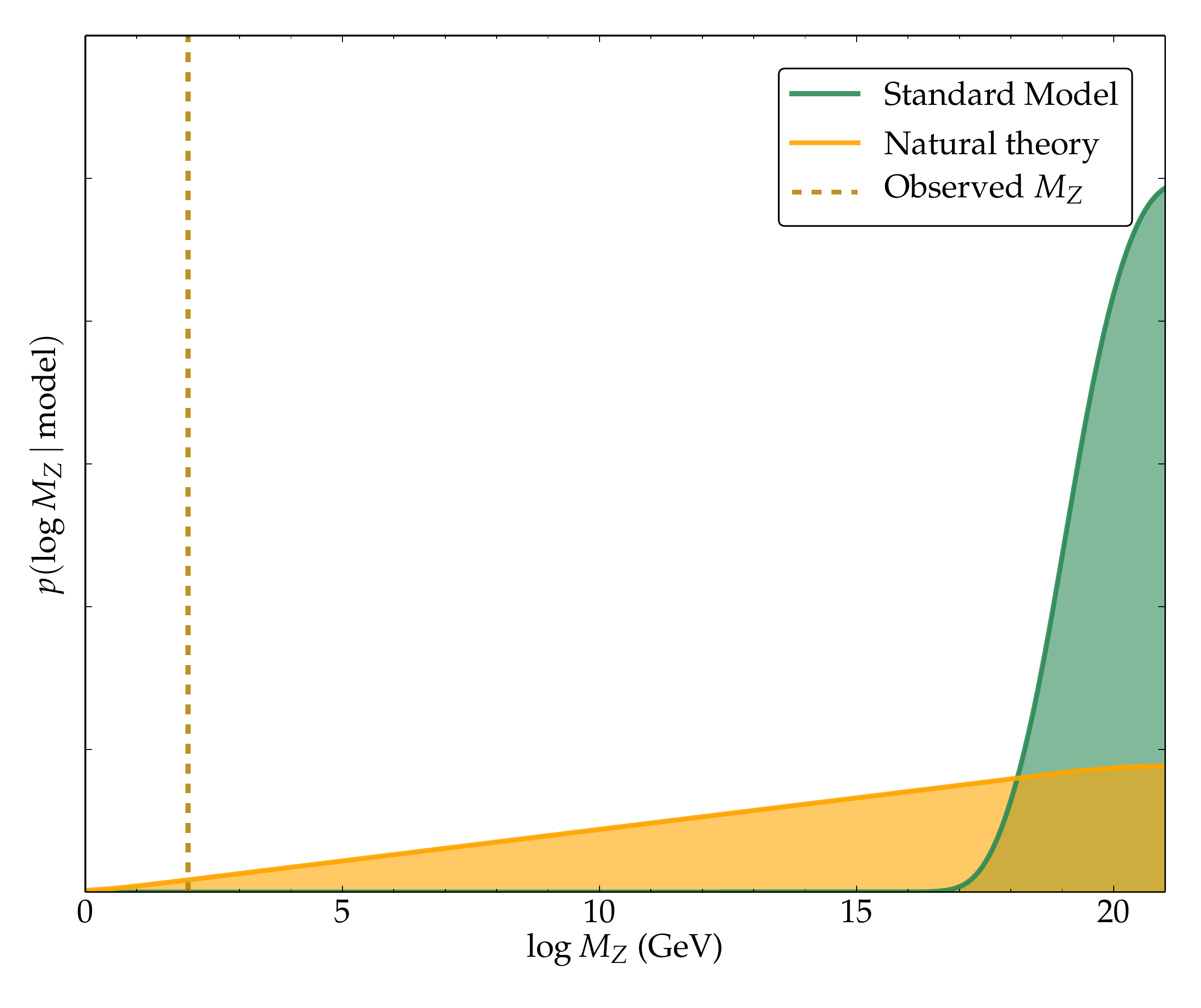}
\caption{Illustration of the evidence, interpreted as a sampling distribution. The SM squanders probability mass at the Planck scale. A natural theory, on the other hand, evades quadratic corrections. The approximately linear behavior in the natural theory results from the fact that the weak scale is the sum of a bare mass and a correction that could be much less than the Planck scale\cite{Fowlie:2014xha}.}
\label{Fig:Evidence}
\end{figure}

The gist of the hierarchy problem and fine-tuning arguments emerge from Bayesian probability --- they are not ingredients or desiderata. To see that the big-hierarchy problem emerges from Bayesian probability, consider the Bayesian evidence (the factor $\pg{\data}{\model}$ in \refeq{Eq:Bayes}) as a function of the data, as plotted in \reffig{Fig:Evidence}. Our data is the observed weak scale, and we consider the SM and a natural theory in which quadratic corrections are truncated before the Planck scale. As a function of the data, the evidence is normalized to unity. Because of quadratic corrections, the SM wastes probability mass near the Planck scale, such that the evidence at the correct weak scale is minuscule --- its generic prediction is that the weak scale is near the Planck scale.\footnote{I assume that the cut-off in the SM is near the Planck scale. If it were higher, the SM's generic prediction for the weak scale would be even worse; if it were lower, it would imply new physics below the Planck scale.}  Natural theories, on the other hand, truncate quadratic corrections at a scale below the Planck scale. Their generic prediction for the weak scale is broad --- the distribution spans the lowest scale to the Planck scale --- but the probability density at the correct weak scale is significantly greater than that in the SM. The LHC constraints slightly weaken our preference for natural theories versus the SM, but the preference for natural theories remains colossal. For example, even in light of LHC results, we ought to place about thirty orders of magnitude more faith in the constrained minimal supersymmetric Standard Model than in the SM\cite{Fowlie:2014xha}.

We calculated the evidences in \reffig{Fig:Evidence} in the usual manner by integrating a likelihood function over a theory's parameter space, denoted by $\theta$, with a suitable measure (the prior distributions):
\begin{align}
\pg{\log\mz}{\model} ={}&  \mz \int\text{d}\theta\pg{\theta}{\model} \times \pg{\mz}{\theta}.
\end{align}
For full details of the calculations in \reffig{Fig:Evidence}, see \refcite{Fowlie:2014xha}. We are not suggesting that our priors correspond to  physical probabilities with which nature picks Lagrangian parameters. Our priors merely reflect our ignorance about which values of the theory's parameters would be realized, were the theory true. Thus, our priors would not be ``wrong'' if we were oblivious to a mechanism in nature that picks Lagrangian parameters such that there is a hierarchy between the weak scale and the Planck scale. If we found such a mechanism, it would be favoured by the Bayesian evidence and solve the hierarchy problem.

With Bayesian probability, the naturalness of a theory is its probability in light of data. Bayesian probability trivializes naturalness arguments --- it is a tautology that natural theories are more probable than unnatural theories.
This is related to Occam's razor \see{sep-ockham}. Bayesian probability justifies Occam's razor and provides an automatic razor \see{Burger} that favors simple theories. Simple theories make clear predictions such that their evidence is sharply peaked; whereas complicated theories make broad predictions such that their evidence is thinly spread. In Bayesian language, naturalness and simplicity are synonymous and the hierarchy problem is a misnomer; there is no problem --- the ``problem'' is simply the observation that theories that generically predict the correct weak scale ought to be favoured.\footnote{There is confusion about whether the weak scale is predictable or calculable in the SM or in a supersymmetric theory. The situation is identical in each theory: if one specifies Lagrangian parameters and a cut-off scale, one can calculate the weak scale. I specify prior distributions for the Lagrangian parameters and cut-off scale and calculate the Bayesian evidence. By ``generic prediction,'' I refer, roughly, to the mode or shape in the Bayesian evidence as a function of the data.} To worry about naturalness is to worry that there might exist a model favoured by the evidence, as in \reffig{Fig:Evidence}.

One might wonder why the weak scale is of special concern; after all, perhaps all measured quantities require fine-tuning. In the SM, for example, we must tune the Yukawa couplings across several orders of magnitude to agree with the measured fermion masses. The answer is that naturalness could concern any measured quantity; the weak scale is special only because the SM's prediction for the weak scale is awful compared to that in, for example, a supersymmetric theory. If a new theory made generic predictions for the Yukawa couplings that resulted in precisely correct fermion masses, it would be favoured by the Bayesian evidence relative to the SM. We do not know of such a theory.

A numerical measure of fine-tuning in supersymmetric theories was developed in the 1980s by Ellis\cite{Ellis:1986yg} and by Barbieri and Giudice\cite{Barbieri:1987fn}. Their measure (henceforth a derivative measure) was based upon derivatives of the weak scale with respect to a theory's fundamental parameters (denoted by $\theta$):
\begin{equation}\label{Eq:BG}
\Delta_i = \frac{\partial \ln\mz}{\partial \ln \theta_i}
\end{equation}
Remarkably, a similar derivative measure results from calculations of the evidence in Bayesian probability \see{Allanach:2007qk,Cabrera:2008tj,Cabrera:2009dm,Fichet:2012sn,Fowlie:2014xha,Fowlie:2014faa}, validating the intuition behind derivative measures. The factor results from a Dirac delta function of the measured weak scale integrated with respect to a parameter, \eg the supersymmetric $\mu$-parameter. If we pick a logarithmic prior for the $\mu$-parameter, we find that
\begin{align}
\begin{split}
\pg{\mz^\prime}{\text{MSSM}} \propto& \int \delta\left(M_Z^\prime - M_Z\right) \p{\mu} \text{d}\mu\\
\propto{}& \int \delta(M_Z^\prime - M_Z) \frac{1}{\mz^\prime} \frac{1}{\Delta_\mu}  \text{d}\mz\\
\propto{}& \frac{1}{\Delta_\mu}.
\end{split}
\end{align}
Note, however, that no derivative measures for parameters other than the chosen parameter (in this case the $\mu$-parameter) emerge in the Bayesian evidence.

By now, however, there are numerous derivative measures \see{Baer:2013gva} --- derivative measures $\Delta_i$ in \refeq{Eq:BG} could be averaged, root-mean-square averaged or minimized, and evaluated at the weak-scale or the high-scale --- built upon a farrago of ideas about naturalness beyond those captured in Bayesian probability. From the Bayesian point of view, those ideas are epistemically irrelevant and specious. This is not a problem or a failure of Bayesian probability to fully account for fine-tuning; our goal is not to emulate foibles of human reasoning about fine-tuning, but to elucidate correct reasoning. Of the miscellany of derivative measures, the best measures are those closest to the spirit and formalism of Bayesian probability.

Ghilencea \etal\cite{Ghilencea:2012gz,Ghilencea:2012qk,Ghilencea:2013nxa,Ghilencea:2013hpa,Antoniadis:2014eta,Ghilencea:2013fka} claim that traditional chi-squared analyses of supersymmetric theories ignore a contribution to the chi-squared from fine-tuning of the electroweak scale. They summarize their work as showing that fine-tuning ``can rule out a model without a detailed chi-squared analysis'' in a frequentist analysis\cite{Ghilencea:2012qk}. The test-statistic in Ghilencea \etal includes the logarithm of the Barbieri-Giudice measure, $\ln\Delta$. Ghilencea \etal approximate the sampling distribution for the $Z$-boson mass with a Dirac delta function --- it is not a random variable and would not vary between repeat experiments. Thus, the quantity $\ln\Delta$  is not a random variable and it makes no sense to claim it could be used as a test-statistic in a frequentist analysis. If the $Z$-boson mass were considered to be a Gaussian random variable, a chi-squared would be a sensible test-statistic, but this chi-squared would be zero because supersymmetric theories are fine-tuned to predict the measured $Z$-boson mass. In other words, contrary to the claims in Ghilencea \etal, there is no fine-tuning penalty in a frequentist analysis. Fine-tuning penalties occur only in Bayesian probability.

\section{The little-hierarchy problem: When should we worry about fine-tuning?}

We stressed that our degree of belief ought to be relative, and that, in Bayesian language, to worry about fine-tuning is to worry that there might exist a theory favoured by the posterior odds.
The fact is that unless a theory exactly reproduces the observed data with no adjustable parameters (I refer to this as a spot-on model), we might always wonder whether a better model exists. That is, there is no concept of ``theory confirmation'' in Bayesian probability --- our favoured theory is only the most probable theory so far, and we might find a better one. In that sense, fine-tuning problems are never resolved. There is no point at which we declare that we have found a natural, untuned theory (unless we find a spot-on model).

When, then, should we worry about fine-tuning? Were we to discover a supersymmetric theory near the weak scale, should we still worry that there could be a more natural explanation for the hierarchy between the weak scale and the Planck scale? To answer this, consider the big-hierarchy problem in the SM. This fine-tuning problem was immediately clear without comparisons to alternative models because one can immediately conceive of ``natural'' theories in which the quadratic corrections are absent. We should always attempt to find more natural theories, but the issue is pressing only if we can readily conceive of a more natural theory. This is not the case with the little-hierarchy problem (or indeed with the cosmological constant\cite{Weinberg:1988cp}); it is a challenge to construct a theory that is favoured by the posterior odds relative to natural theories that suffer from a little-hierarchy problem. Admittedly, this is somewhat oxymoronic --- we should only worry about fine-tuning problems if we can envisage solutions. The point is, however, that fine-tuning problems are not problems; they are simply the observation that a more probable theory might exist. If we struggle to construct such a theory, there is no immediate problem, though if in the future we find one, it would be favoured.

If we discovered evidence for a natural theory slightly above the weak scale, at $10\tev$, for example, there might be puzzlement about why it was at $10\tev$ rather than $1\tev$; certainly some might ask, why is there a little-hierarchy? Why is nature described by a natural theory at $10\tev$ rather than a natural theory at $1\tev$? After all, would not a natural theory at $1\tev$ be most natural? Our answer to the latter question is negative --- naturalness is nothing but Bayesian evidence. The evidence for a theory that predicts new physics at $1\tev$ would be minuscule if we saw new physics only at $10\tev$. The former question, on the other hand, is purely metaphysical; it cannot be posed in Bayesian language.


\section{The Standard Model is not an effective theory}

Let us turn to the first response to the little-hierarchy problem outlined in the introduction --- that the SM is not an effective theory. In this case, presumably, finite renormalized parameters in the SM, in a particular renormalization scheme, are fundamental parameters. This is at odds with the modern understanding of renormalization; in this approach, renormalization is viewed as it was prior to Wilson's insights\cite{Wilson:1973jj} --- an algorithm for ``sweeping the infinities under the rug.'' What are sensible priors for the theory's parameters? What, then, is the generic prediction of such a model for the weak scale? This is problematic; we claim that this is a fundamental theory, closing the door on any mechanism that could determine the parameters, indicate that dimensionless parameters might be close to one, or bound the parameters. The priors for the parameters in such a theory and its prediction for the weak scale are ill-defined improper distributions --- infinitesimal over an infinite range.\footnote{This argument fails for quantum field theories with no adjustable parameters or a single adjustable parameter, such as QCD, because a single parameter is an arbitrary definition of a unit of mass.} This prediction for the weak scale is much worse than that of the SM in \reffig{Fig:Evidence}. Admittedly, such a model may replicate our observed data, but even the SM is a more probable explanation of that data. If the SM is interpreted as an effective theory, it is reasonable that our prior distributions are proper distributions, that is, that our priors reach zero asymptotically, because a mechanism in the ultra-violet theory could determine or bound the parameters in the SM. This  disfavours scales that far exceed the cut-off scale or couplings that far exceed unity.

\section{Technical naturalness}

Before summarizing, I briefly turn to ``technical naturalness'' --- a notion of naturalness related to big and small numbers espoused by Dirac in the 1930s\cite{Dirac:1937ti}. Dirac frowned upon any big or small fundamental parameters in a theory. From a Bayesian perspective, Dirac's preference for numbers of order unity might be dismissed as an eccentric, subjective prior. Our priors ought to reflect our state of knowledge; we must ask ourselves whether there is sufficient reason for believing that small numbers are improbable in nature. It is plausible, but hardly compelling and possibly circular, that if our theory's parameters were exact solutions of an unknown fundamental theory, exact solutions of order one might be more probable.

Dirac's idea was modernized by 't Hooft in the 1970s\cite{'tHooft:1979bh}. Unlike Dirac, 't Hooft tolerated small numbers, but only if they were connected to a symmetry:
\begin{iquote}
at any scale $\mu$, a physical parameter or set of physical parameters $\alpha_i(\mu)$ is allowed to be very small only if the replacement $\alpha_i(\mu)=0$ would increase the symmetry of the system.
\end{iquote}
't Hooft's modification could be justified by Bayesian probability. Because such a small parameter could originate from the microscopic details of a mechanism of symmetry breaking, we might find it, a priori, more plausible that it is small.

In \refsec{Sec:BayesNat}, naturalness was favoured in the evidence by a combination of prior beliefs and experimental data in the likelihood. In contrast, technical naturalness only concerns appropriate prior beliefs in a theory's parameters. Dirac's and 't Hooft's ideas about technical naturalness hinge upon whether it is appropriate to penalize big and small numbers in our priors. Although such penalties could be justified in particular theories, there is no reason why we ought to pick priors that favour technical naturalness in all theories in high-energy physics.

\section{Summary and discussion}

In light of data from LEP and the LHC, many are questioning the principle of naturalness and the significance of the hierarchy problem --- perhaps naturalness was an ill-defined prejudiced basis for scientific knowledge? We have argued, however, that naturalness is not an aesthetic principle. Naturalness is grounded in Bayesian probability, which is arguably the unique framework for updating our opinions in light of experimental data, that is, the unique framework for scientific inference. Because Bayesian probability is a mathematical framework rather than a scientific theory, it makes no sense to argue that it has been ``falsified'' because we observed ``improbable'' data. As a helpful analogy, suppose I claimed to have falsified probability theory because I observed a coin land on heads 100 times in a row, which I deemed improbable. Besides my reasoning being somewhat circular, my observations could only discriminate between rival explanations for the behaviour of the coin. I might infer that the coin is biased, but I cannot conclude that I must reject probability theory. On the other hand, we found limited support for technical naturalness.

The big-hierarchy problem is not in itself a problem; it is merely the fact that compared with supersymmetric or technicolor theories, the SM makes an awful prediction for the weak scale. The big-hierarchy problem in the SM is immediately clear without comparisons to alternative models because one can immediately conceive of ``natural'' theories in which the quadratic corrections are absent. In light of $M_Z/M_P\lesssim 10^{-16}$, we should doubt the SM. In contrast, there is no reason to doubt supersymmetric or natural theories because of the little-hierarchy problem --- there is no alternative explanation under which the data is more probable. We cannot consider whether a supersymmetric or natural theory is probable --- we must consider whether it is more or less probable than alternative explanations, such as the SM.

We found that, from a Bayesian perspective, there is no support for the argument that the SM is a fundamental theory. This exacerbates fine-tuning problems because whereas the SM makes an awful generic prediction that the weak scale is close to the Planck scale, such a theory makes no generic predictions for the weak scale.

In summary, the little-hierarchy problem is a little problem; we cannot reject the concept of naturalness because of the absence of new physics near the weak scale. This faulty reasoning stems from a misunderstanding of the foundations of naturalness arguments. As frustrating as it is, until we find it or reach the highest energy scales, the most probable theories, such as supersymmetric theories, predict new physics that is always just around the corner and we should remain positive about the prospects for new physics in the second run of the LHC.

\bibliography{hierarchy}

\begin{thebibliography}{75}%
\makeatletter
\providecommand \@ifxundefined [1]{%
 \@ifx{#1\undefined}
}%
\providecommand \@ifnum [1]{%
 \ifnum #1\expandafter \@firstoftwo
 \else \expandafter \@secondoftwo
 \fi
}%
\providecommand \@ifx [1]{%
 \ifx #1\expandafter \@firstoftwo
 \else \expandafter \@secondoftwo
 \fi
}%
\providecommand \natexlab [1]{#1}%
\providecommand \enquote  [1]{``#1''}%
\providecommand \bibnamefont  [1]{#1}%
\providecommand \bibfnamefont [1]{#1}%
\providecommand \citenamefont [1]{#1}%
\providecommand \href@noop [0]{\@secondoftwo}%
\providecommand \href [0]{\begingroup \@sanitize@url \@href}%
\providecommand \@href[1]{\@@startlink{#1}\@@href}%
\providecommand \@@href[1]{\endgroup#1\@@endlink}%
\providecommand \@sanitize@url [0]{\catcode `\\12\catcode `\$12\catcode
  `\&12\catcode `\#12\catcode `\^12\catcode `\_12\catcode `\%12\relax}%
\providecommand \@@startlink[1]{}%
\providecommand \@@endlink[0]{}%
\providecommand \url  [0]{\begingroup\@sanitize@url \@url }%
\providecommand \@url [1]{\endgroup\@href {#1}{\urlprefix }}%
\providecommand \urlprefix  [0]{URL }%
\providecommand \Eprint [0]{\href }%
\providecommand \doibase [0]{http://dx.doi.org/}%
\providecommand \selectlanguage [0]{\@gobble}%
\providecommand \bibinfo  [0]{\@secondoftwo}%
\providecommand \bibfield  [0]{\@secondoftwo}%
\providecommand \translation [1]{[#1]}%
\providecommand \BibitemOpen [0]{}%
\providecommand \bibitemStop [0]{}%
\providecommand \bibitemNoStop [0]{.\EOS\space}%
\providecommand \EOS [0]{\spacefactor3000\relax}%
\providecommand \BibitemShut  [1]{\csname bibitem#1\endcsname}%
\let\auto@bib@innerbib\@empty
\bibitem [{\citenamefont {Nath}\ \emph {et~al.}(2010)\citenamefont {Nath},
  \citenamefont {Nelson}, \citenamefont {Davoudiasl}, \citenamefont {Dutta},
  \citenamefont {Feldman} \emph {et~al.}}]{Nath:2010zj}%
  \BibitemOpen
  \bibfield  {author} {\bibinfo {author} {\bibfnamefont {P.}~\bibnamefont
  {Nath}}, \bibinfo {author} {\bibfnamefont {B.~D.}\ \bibnamefont {Nelson}},
  \bibinfo {author} {\bibfnamefont {H.}~\bibnamefont {Davoudiasl}}, \bibinfo
  {author} {\bibfnamefont {B.}~\bibnamefont {Dutta}}, \bibinfo {author}
  {\bibfnamefont {D.}~\bibnamefont {Feldman}},  \emph {et~al.},\ }\href
  {\doibase 10.1016/j.nuclphysbps.2010.03.001} {\bibfield  {journal} {\bibinfo
  {journal} {Nucl.Phys.Proc.Suppl.}\ }\textbf {\bibinfo {volume} {200-202}},\
  \bibinfo {pages} {185} (\bibinfo {year} {2010})},\ \Eprint
  {http://arxiv.org/abs/1001.2693} {arXiv:1001.2693 [hep-ph]} \BibitemShut
  {NoStop}%
\bibitem [{\citenamefont {Kane}\ and\ \citenamefont
  {Pierce}(2008)}]{Kane:2008zz}%
  \BibitemOpen
  \bibfield  {author} {\bibinfo {author} {\bibfnamefont {G.}~\bibnamefont
  {Kane}}\ and\ \bibinfo {author} {\bibfnamefont {A.}~\bibnamefont {Pierce}},\
  }\href@noop {} {\emph {\bibinfo {title} {{Perspectives on LHC physics}}}}\
  (\bibinfo  {publisher} {World Scientific},\ \bibinfo {year}
  {2008})\BibitemShut {NoStop}%
\bibitem [{\citenamefont {Pich}(2015)}]{Pich:2015tqa}%
  \BibitemOpen
  \bibfield  {author} {\bibinfo {author} {\bibfnamefont {A.}~\bibnamefont
  {Pich}},\ }\href@noop {} {\bibfield  {journal} {\bibinfo  {journal} {CNUM:
  C14-07-02}\ } (\bibinfo {year} {2015})},\ \Eprint
  {http://arxiv.org/abs/1505.01813} {arXiv:1505.01813 [hep-ph]} \BibitemShut
  {NoStop}%
\bibitem [{\citenamefont {Strumia}(2015)}]{Strumia:2015dca}%
  \BibitemOpen
  \bibfield  {author} {\bibinfo {author} {\bibfnamefont {A.}~\bibnamefont
  {Strumia}},\ }\href@noop {} {\bibfield  {journal} {\bibinfo  {journal} {CNUM:
  C15-03-14}\ } (\bibinfo {year} {2015})},\ \Eprint
  {http://arxiv.org/abs/1504.08331} {arXiv:1504.08331 [hep-ph]} \BibitemShut
  {NoStop}%
\bibitem [{\citenamefont {Ellis}(2015)}]{Ellis:2014cga}%
  \BibitemOpen
  \bibfield  {author} {\bibinfo {author} {\bibfnamefont {J.}~\bibnamefont
  {Ellis}},\ }\href@noop {} {\bibfield  {journal} {\bibinfo  {journal} {PoS}\
  }\textbf {\bibinfo {volume} {Beauty2014}},\ \bibinfo {pages} {056} (\bibinfo
  {year} {2015})},\ \Eprint {http://arxiv.org/abs/1412.2666} {arXiv:1412.2666
  [hep-ph]} \BibitemShut {NoStop}%
\bibitem [{\citenamefont {Baer}\ \emph {et~al.}(2015)\citenamefont {Baer},
  \citenamefont {Barger},\ and\ \citenamefont {Savoy}}]{Baer:2015fsa}%
  \BibitemOpen
  \bibfield  {author} {\bibinfo {author} {\bibfnamefont {H.}~\bibnamefont
  {Baer}}, \bibinfo {author} {\bibfnamefont {V.}~\bibnamefont {Barger}}, \ and\
  \bibinfo {author} {\bibfnamefont {M.}~\bibnamefont {Savoy}},\ }\href
  {\doibase 10.1088/0031-8949/90/6/068003} {\bibfield  {journal} {\bibinfo
  {journal} {Phys.Scripta}\ }\textbf {\bibinfo {volume} {90}},\ \bibinfo
  {pages} {068003} (\bibinfo {year} {2015})},\ \Eprint
  {http://arxiv.org/abs/1502.04127} {arXiv:1502.04127 [hep-ph]} \BibitemShut
  {NoStop}%
\bibitem [{\citenamefont {Weinberg}(1976)}]{Weinberg:1975gm}%
  \BibitemOpen
  \bibfield  {author} {\bibinfo {author} {\bibfnamefont {S.}~\bibnamefont
  {Weinberg}},\ }\href {\doibase 10.1103/PhysRevD.13.974} {\bibfield  {journal}
  {\bibinfo  {journal} {Phys.Rev.}\ }\textbf {\bibinfo {volume} {D13}},\
  \bibinfo {pages} {974} (\bibinfo {year} {1976})}\BibitemShut {NoStop}%
\bibitem [{\citenamefont {Weinberg}(1979)}]{Weinberg:1979bn}%
  \BibitemOpen
  \bibfield  {author} {\bibinfo {author} {\bibfnamefont {S.}~\bibnamefont
  {Weinberg}},\ }\href {\doibase 10.1103/PhysRevD.19.1277} {\bibfield
  {journal} {\bibinfo  {journal} {Phys.Rev.}\ }\textbf {\bibinfo {volume}
  {D19}},\ \bibinfo {pages} {1277} (\bibinfo {year} {1979})}\BibitemShut
  {NoStop}%
\bibitem [{\citenamefont {Susskind}(1979)}]{Susskind:1978ms}%
  \BibitemOpen
  \bibfield  {author} {\bibinfo {author} {\bibfnamefont {L.}~\bibnamefont
  {Susskind}},\ }\href {\doibase 10.1103/PhysRevD.20.2619} {\bibfield
  {journal} {\bibinfo  {journal} {Phys.Rev.}\ }\textbf {\bibinfo {volume}
  {D20}},\ \bibinfo {pages} {2619} (\bibinfo {year} {1979})}\BibitemShut
  {NoStop}%
\bibitem [{\citenamefont {Gildener}(1976)}]{Gildener:1976ai}%
  \BibitemOpen
  \bibfield  {author} {\bibinfo {author} {\bibfnamefont {E.}~\bibnamefont
  {Gildener}},\ }\href {\doibase 10.1103/PhysRevD.14.1667} {\bibfield
  {journal} {\bibinfo  {journal} {Phys.Rev.}\ }\textbf {\bibinfo {volume}
  {D14}},\ \bibinfo {pages} {1667} (\bibinfo {year} {1976})}\BibitemShut
  {NoStop}%
\bibitem [{\citenamefont {Glashow}(1961)}]{Glashow:1961tr}%
  \BibitemOpen
  \bibfield  {author} {\bibinfo {author} {\bibfnamefont {S.}~\bibnamefont
  {Glashow}},\ }\href {\doibase 10.1016/0029-5582(61)90469-2} {\bibfield
  {journal} {\bibinfo  {journal} {Nucl.Phys.}\ }\textbf {\bibinfo {volume}
  {22}},\ \bibinfo {pages} {579} (\bibinfo {year} {1961})}\BibitemShut
  {NoStop}%
\bibitem [{\citenamefont {Salam}(1968)}]{Salam:1968rm}%
  \BibitemOpen
  \bibfield  {author} {\bibinfo {author} {\bibfnamefont {A.}~\bibnamefont
  {Salam}},\ }\href@noop {} {\bibfield  {journal} {\bibinfo  {journal}
  {Conf.Proc.}\ }\textbf {\bibinfo {volume} {C680519}},\ \bibinfo {pages} {367}
  (\bibinfo {year} {1968})}\BibitemShut {NoStop}%
\bibitem [{\citenamefont {Weinberg}(1967)}]{Weinberg:1967tq}%
  \BibitemOpen
  \bibfield  {author} {\bibinfo {author} {\bibfnamefont {S.}~\bibnamefont
  {Weinberg}},\ }\href {\doibase 10.1103/PhysRevLett.19.1264} {\bibfield
  {journal} {\bibinfo  {journal} {Phys.Rev.Lett.}\ }\textbf {\bibinfo {volume}
  {19}},\ \bibinfo {pages} {1264} (\bibinfo {year} {1967})}\BibitemShut
  {NoStop}%
\bibitem [{\citenamefont {Salam}\ and\ \citenamefont
  {Strathdee}(1974)}]{Salam:1974yz}%
  \BibitemOpen
  \bibfield  {author} {\bibinfo {author} {\bibfnamefont {A.}~\bibnamefont
  {Salam}}\ and\ \bibinfo {author} {\bibfnamefont {J.}~\bibnamefont
  {Strathdee}},\ }\href {\doibase 10.1016/0550-3213(74)90537-9} {\bibfield
  {journal} {\bibinfo  {journal} {Nucl.Phys.}\ }\textbf {\bibinfo {volume}
  {B76}},\ \bibinfo {pages} {477} (\bibinfo {year} {1974})}\BibitemShut
  {NoStop}%
\bibitem [{\citenamefont {Haber}\ and\ \citenamefont
  {Kane}(1985)}]{Haber:1984rc}%
  \BibitemOpen
  \bibfield  {author} {\bibinfo {author} {\bibfnamefont {H.~E.}\ \bibnamefont
  {Haber}}\ and\ \bibinfo {author} {\bibfnamefont {G.~L.}\ \bibnamefont
  {Kane}},\ }\href {\doibase 10.1016/0370-1573(85)90051-1} {\bibfield
  {journal} {\bibinfo  {journal} {Phys.Rept.}\ }\textbf {\bibinfo {volume}
  {117}},\ \bibinfo {pages} {75} (\bibinfo {year} {1985})}\BibitemShut
  {NoStop}%
\bibitem [{\citenamefont {Nilles}(1984)}]{Nilles:1983ge}%
  \BibitemOpen
  \bibfield  {author} {\bibinfo {author} {\bibfnamefont {H.~P.}\ \bibnamefont
  {Nilles}},\ }\href {\doibase 10.1016/0370-1573(84)90008-5} {\bibfield
  {journal} {\bibinfo  {journal} {Phys.Rept.}\ }\textbf {\bibinfo {volume}
  {110}},\ \bibinfo {pages} {1} (\bibinfo {year} {1984})}\BibitemShut {NoStop}%
\bibitem [{\citenamefont {Witten}(1981)}]{Witten:1981nf}%
  \BibitemOpen
  \bibfield  {author} {\bibinfo {author} {\bibfnamefont {E.}~\bibnamefont
  {Witten}},\ }\href {\doibase 10.1016/0550-3213(81)90006-7} {\bibfield
  {journal} {\bibinfo  {journal} {Nucl.Phys.}\ }\textbf {\bibinfo {volume}
  {B188}},\ \bibinfo {pages} {513} (\bibinfo {year} {1981})}\BibitemShut
  {NoStop}%
\bibitem [{\citenamefont {Jackiw}\ and\ \citenamefont
  {Johnson}(1973)}]{Jackiw:1973tr}%
  \BibitemOpen
  \bibfield  {author} {\bibinfo {author} {\bibfnamefont {R.}~\bibnamefont
  {Jackiw}}\ and\ \bibinfo {author} {\bibfnamefont {K.}~\bibnamefont
  {Johnson}},\ }\href {\doibase 10.1103/PhysRevD.8.2386} {\bibfield  {journal}
  {\bibinfo  {journal} {Phys.Rev.}\ }\textbf {\bibinfo {volume} {D8}},\
  \bibinfo {pages} {2386} (\bibinfo {year} {1973})}\BibitemShut {NoStop}%
\bibitem [{\citenamefont {Cornwall}\ and\ \citenamefont
  {Norton}(1973)}]{Cornwall:1973ts}%
  \BibitemOpen
  \bibfield  {author} {\bibinfo {author} {\bibfnamefont {J.}~\bibnamefont
  {Cornwall}}\ and\ \bibinfo {author} {\bibfnamefont {R.}~\bibnamefont
  {Norton}},\ }\href {\doibase 10.1103/PhysRevD.8.3338} {\bibfield  {journal}
  {\bibinfo  {journal} {Phys.Rev.}\ }\textbf {\bibinfo {volume} {D8}},\
  \bibinfo {pages} {3338} (\bibinfo {year} {1973})}\BibitemShut {NoStop}%
\bibitem [{\citenamefont {Eichten}\ and\ \citenamefont
  {Feinberg}(1974)}]{Eichten:1974et}%
  \BibitemOpen
  \bibfield  {author} {\bibinfo {author} {\bibfnamefont {E.}~\bibnamefont
  {Eichten}}\ and\ \bibinfo {author} {\bibfnamefont {F.}~\bibnamefont
  {Feinberg}},\ }\href {\doibase 10.1103/PhysRevD.10.3254} {\bibfield
  {journal} {\bibinfo  {journal} {Phys.Rev.}\ }\textbf {\bibinfo {volume}
  {D10}},\ \bibinfo {pages} {3254} (\bibinfo {year} {1974})}\BibitemShut
  {NoStop}%
\bibitem [{\citenamefont {Arkani-Hamed}\ \emph {et~al.}(1999)\citenamefont
  {Arkani-Hamed}, \citenamefont {Dimopoulos},\ and\ \citenamefont
  {Dvali}}]{ArkaniHamed:1998nn}%
  \BibitemOpen
  \bibfield  {author} {\bibinfo {author} {\bibfnamefont {N.}~\bibnamefont
  {Arkani-Hamed}}, \bibinfo {author} {\bibfnamefont {S.}~\bibnamefont
  {Dimopoulos}}, \ and\ \bibinfo {author} {\bibfnamefont {G.}~\bibnamefont
  {Dvali}},\ }\href {\doibase 10.1103/PhysRevD.59.086004} {\bibfield  {journal}
  {\bibinfo  {journal} {Phys.Rev.}\ }\textbf {\bibinfo {volume} {D59}},\
  \bibinfo {pages} {086004} (\bibinfo {year} {1999})},\ \Eprint
  {http://arxiv.org/abs/hep-ph/9807344} {arXiv:hep-ph/9807344 [hep-ph]}
  \BibitemShut {NoStop}%
\bibitem [{\citenamefont {Arkani-Hamed}\ \emph {et~al.}(1998)\citenamefont
  {Arkani-Hamed}, \citenamefont {Dimopoulos},\ and\ \citenamefont
  {Dvali}}]{ArkaniHamed:1998rs}%
  \BibitemOpen
  \bibfield  {author} {\bibinfo {author} {\bibfnamefont {N.}~\bibnamefont
  {Arkani-Hamed}}, \bibinfo {author} {\bibfnamefont {S.}~\bibnamefont
  {Dimopoulos}}, \ and\ \bibinfo {author} {\bibfnamefont {G.}~\bibnamefont
  {Dvali}},\ }\href {\doibase 10.1016/S0370-2693(98)00466-3} {\bibfield
  {journal} {\bibinfo  {journal} {Phys.Lett.}\ }\textbf {\bibinfo {volume}
  {B429}},\ \bibinfo {pages} {263} (\bibinfo {year} {1998})},\ \Eprint
  {http://arxiv.org/abs/9803315} {arXiv:9803315 [hep-ph]} \BibitemShut
  {NoStop}%
\bibitem [{\citenamefont {Graham}\ \emph {et~al.}(2015)\citenamefont {Graham},
  \citenamefont {Kaplan},\ and\ \citenamefont {Rajendran}}]{Graham:2015cka}%
  \BibitemOpen
  \bibfield  {author} {\bibinfo {author} {\bibfnamefont {P.~W.}\ \bibnamefont
  {Graham}}, \bibinfo {author} {\bibfnamefont {D.~E.}\ \bibnamefont {Kaplan}},
  \ and\ \bibinfo {author} {\bibfnamefont {S.}~\bibnamefont {Rajendran}},\
  }\href@noop {} {\  (\bibinfo {year} {2015})},\ \Eprint
  {http://arxiv.org/abs/1504.07551} {arXiv:1504.07551 [hep-ph]} \BibitemShut
  {NoStop}%
\bibitem [{\citenamefont {Assmann}\ \emph {et~al.}(2002)\citenamefont
  {Assmann}, \citenamefont {Lamont},\ and\ \citenamefont
  {Myers}}]{Assmann:2002th}%
  \BibitemOpen
  \bibfield  {author} {\bibinfo {author} {\bibfnamefont {R.}~\bibnamefont
  {Assmann}}, \bibinfo {author} {\bibfnamefont {M.}~\bibnamefont {Lamont}}, \
  and\ \bibinfo {author} {\bibfnamefont {S.}~\bibnamefont {Myers}},\ }\href
  {\doibase 10.1016/S0920-5632(02)90005-8} {\bibfield  {journal} {\bibinfo
  {journal} {Nucl.Phys.Proc.Suppl.}\ }\textbf {\bibinfo {volume} {109B}},\
  \bibinfo {pages} {17} (\bibinfo {year} {2002})}\BibitemShut {NoStop}%
\bibitem [{\citenamefont {Barate}\ \emph {et~al.}(2003)\citenamefont {Barate}
  \emph {et~al.}}]{Barate:2003sz}%
  \BibitemOpen
  \bibfield  {author} {\bibinfo {author} {\bibfnamefont {R.}~\bibnamefont
  {Barate}} \emph {et~al.} (\bibinfo {collaboration} {LEP Working Group for
  Higgs boson searches, ALEPH, DELPHI, L3, OPAL}),\ }\href {\doibase
  10.1016/S0370-2693(03)00614-2} {\bibfield  {journal} {\bibinfo  {journal}
  {Phys.Lett.}\ }\textbf {\bibinfo {volume} {B565}},\ \bibinfo {pages} {61}
  (\bibinfo {year} {2003})},\ \Eprint {http://arxiv.org/abs/hep-ex/0306033}
  {arXiv:hep-ex/0306033 [hep-ex]} \BibitemShut {NoStop}%
\bibitem [{\citenamefont {Barbieri}\ and\ \citenamefont
  {Strumia}(2000)}]{Barbieri:2000gf}%
  \BibitemOpen
  \bibfield  {author} {\bibinfo {author} {\bibfnamefont {R.}~\bibnamefont
  {Barbieri}}\ and\ \bibinfo {author} {\bibfnamefont {A.}~\bibnamefont
  {Strumia}},\ }\href@noop {} {\bibfield  {journal} {\bibinfo  {journal} {CNUM:
  C00-07-19}\ } (\bibinfo {year} {2000})},\ \Eprint
  {http://arxiv.org/abs/hep-ph/0007265} {arXiv:hep-ph/0007265 [hep-ph]}
  \BibitemShut {NoStop}%
\bibitem [{\citenamefont {Bechtle}\ \emph {et~al.}(2015)\citenamefont
  {Bechtle}, \citenamefont {Plehn},\ and\ \citenamefont
  {Sander}}]{Bechtle:2015nta}%
  \BibitemOpen
  \bibfield  {author} {\bibinfo {author} {\bibfnamefont {P.}~\bibnamefont
  {Bechtle}}, \bibinfo {author} {\bibfnamefont {T.}~\bibnamefont {Plehn}}, \
  and\ \bibinfo {author} {\bibfnamefont {C.}~\bibnamefont {Sander}},\ }\enquote
  {\bibinfo {title} {{The Large Hadron Collider: Harvest of Run 1}},}\ \
  (\bibinfo  {publisher} {Springer},\ \bibinfo {year} {2015})\ Chap.\ \bibinfo
  {chapter} {The Status of Supersymmetry after the LHC Run 1},\ \Eprint
  {http://arxiv.org/abs/1506.03091} {arXiv:1506.03091 [hep-ex]} \BibitemShut
  {NoStop}%
\bibitem [{\citenamefont {Strumia}(2011)}]{Strumia:2011dv}%
  \BibitemOpen
  \bibfield  {author} {\bibinfo {author} {\bibfnamefont {A.}~\bibnamefont
  {Strumia}},\ }\href {\doibase 10.1007/JHEP04(2011)073} {\bibfield  {journal}
  {\bibinfo  {journal} {JHEP}\ }\textbf {\bibinfo {volume} {1104}},\ \bibinfo
  {pages} {073} (\bibinfo {year} {2011})},\ \Eprint
  {http://arxiv.org/abs/1101.2195} {arXiv:1101.2195 [hep-ph]} \BibitemShut
  {NoStop}%
\bibitem [{\citenamefont {Aad}\ \emph {et~al.}(2012)\citenamefont {Aad} \emph
  {et~al.}}]{Aad:2012tfa}%
  \BibitemOpen
  \bibfield  {author} {\bibinfo {author} {\bibfnamefont {G.}~\bibnamefont
  {Aad}} \emph {et~al.} (\bibinfo {collaboration} {ATLAS}),\ }\href {\doibase
  10.1016/j.physletb.2012.08.020} {\bibfield  {journal} {\bibinfo  {journal}
  {Phys.Lett.}\ }\textbf {\bibinfo {volume} {B716}},\ \bibinfo {pages} {1}
  (\bibinfo {year} {2012})},\ \Eprint {http://arxiv.org/abs/1207.7214}
  {arXiv:1207.7214 [hep-ex]} \BibitemShut {NoStop}%
\bibitem [{\citenamefont {Chatrchyan}\ \emph {et~al.}(2012)\citenamefont
  {Chatrchyan} \emph {et~al.}}]{Chatrchyan:2012ufa}%
  \BibitemOpen
  \bibfield  {author} {\bibinfo {author} {\bibfnamefont {S.}~\bibnamefont
  {Chatrchyan}} \emph {et~al.} (\bibinfo {collaboration} {CMS}),\ }\href
  {\doibase 10.1016/j.physletb.2012.08.021} {\bibfield  {journal} {\bibinfo
  {journal} {Phys.Lett.}\ }\textbf {\bibinfo {volume} {B716}},\ \bibinfo
  {pages} {30} (\bibinfo {year} {2012})},\ \Eprint
  {http://arxiv.org/abs/1207.7235} {arXiv:1207.7235 [hep-ex]} \BibitemShut
  {NoStop}%
\bibitem [{\citenamefont {Dine}(2015)}]{Dine:2015xga}%
  \BibitemOpen
  \bibfield  {author} {\bibinfo {author} {\bibfnamefont {M.}~\bibnamefont
  {Dine}},\ }\href@noop {} {\  (\bibinfo {year} {2015})},\ \Eprint
  {http://arxiv.org/abs/1501.01035} {arXiv:1501.01035 [hep-ph]} \BibitemShut
  {NoStop}%
\bibitem [{\citenamefont {Salvio}\ and\ \citenamefont
  {Strumia}(2014)}]{Salvio:2014soa}%
  \BibitemOpen
  \bibfield  {author} {\bibinfo {author} {\bibfnamefont {A.}~\bibnamefont
  {Salvio}}\ and\ \bibinfo {author} {\bibfnamefont {A.}~\bibnamefont
  {Strumia}},\ }\href {\doibase 10.1007/JHEP06(2014)080} {\bibfield  {journal}
  {\bibinfo  {journal} {JHEP}\ }\textbf {\bibinfo {volume} {1406}},\ \bibinfo
  {pages} {080} (\bibinfo {year} {2014})},\ \Eprint
  {http://arxiv.org/abs/1403.4226} {arXiv:1403.4226 [hep-ph]} \BibitemShut
  {NoStop}%
\bibitem [{\citenamefont {Giudice}\ \emph {et~al.}(2015)\citenamefont
  {Giudice}, \citenamefont {Isidori}, \citenamefont {Salvio},\ and\
  \citenamefont {Strumia}}]{Giudice:2014tma}%
  \BibitemOpen
  \bibfield  {author} {\bibinfo {author} {\bibfnamefont {G.~F.}\ \bibnamefont
  {Giudice}}, \bibinfo {author} {\bibfnamefont {G.}~\bibnamefont {Isidori}},
  \bibinfo {author} {\bibfnamefont {A.}~\bibnamefont {Salvio}}, \ and\ \bibinfo
  {author} {\bibfnamefont {A.}~\bibnamefont {Strumia}},\ }\href {\doibase
  10.1007/JHEP02(2015)137} {\bibfield  {journal} {\bibinfo  {journal} {JHEP}\
  }\textbf {\bibinfo {volume} {1502}},\ \bibinfo {pages} {137} (\bibinfo {year}
  {2015})},\ \Eprint {http://arxiv.org/abs/1412.2769} {arXiv:1412.2769
  [hep-ph]} \BibitemShut {NoStop}%
\bibitem [{\citenamefont {Bardeen}(1995)}]{Bardeen:1995kv}%
  \BibitemOpen
  \bibfield  {author} {\bibinfo {author} {\bibfnamefont {W.~A.}\ \bibnamefont
  {Bardeen}},\ }\href@noop {} {\bibfield  {journal} {\bibinfo  {journal} {CNUM:
  C95-08-27.3}\ } (\bibinfo {year} {1995})}\BibitemShut {NoStop}%
\bibitem [{\citenamefont {Heikinheimo}\ \emph {et~al.}(2014)\citenamefont
  {Heikinheimo}, \citenamefont {Racioppi}, \citenamefont {Raidal},
  \citenamefont {Spethmann},\ and\ \citenamefont
  {Tuominen}}]{Heikinheimo:2013fta}%
  \BibitemOpen
  \bibfield  {author} {\bibinfo {author} {\bibfnamefont {M.}~\bibnamefont
  {Heikinheimo}}, \bibinfo {author} {\bibfnamefont {A.}~\bibnamefont
  {Racioppi}}, \bibinfo {author} {\bibfnamefont {M.}~\bibnamefont {Raidal}},
  \bibinfo {author} {\bibfnamefont {C.}~\bibnamefont {Spethmann}}, \ and\
  \bibinfo {author} {\bibfnamefont {K.}~\bibnamefont {Tuominen}},\ }\href
  {\doibase 10.1142/S0217732314500771} {\bibfield  {journal} {\bibinfo
  {journal} {Mod.Phys.Lett.}\ }\textbf {\bibinfo {volume} {A29}},\ \bibinfo
  {pages} {1450077} (\bibinfo {year} {2014})},\ \Eprint
  {http://arxiv.org/abs/1304.7006} {arXiv:1304.7006 [hep-ph]} \BibitemShut
  {NoStop}%
\bibitem [{\citenamefont {Kannike}\ \emph {et~al.}(2014)\citenamefont
  {Kannike}, \citenamefont {Racioppi},\ and\ \citenamefont
  {Raidal}}]{Kannike:2014mia}%
  \BibitemOpen
  \bibfield  {author} {\bibinfo {author} {\bibfnamefont {K.}~\bibnamefont
  {Kannike}}, \bibinfo {author} {\bibfnamefont {A.}~\bibnamefont {Racioppi}}, \
  and\ \bibinfo {author} {\bibfnamefont {M.}~\bibnamefont {Raidal}},\ }\href
  {\doibase 10.1007/JHEP06(2014)154} {\bibfield  {journal} {\bibinfo  {journal}
  {JHEP}\ }\textbf {\bibinfo {volume} {1406}},\ \bibinfo {pages} {154}
  (\bibinfo {year} {2014})},\ \Eprint {http://arxiv.org/abs/1405.3987}
  {arXiv:1405.3987 [hep-ph]} \BibitemShut {NoStop}%
\bibitem [{\citenamefont {Gabrielli}\ \emph {et~al.}(2014)\citenamefont
  {Gabrielli}, \citenamefont {Heikinheimo}, \citenamefont {Kannike},
  \citenamefont {Racioppi}, \citenamefont {Raidal} \emph
  {et~al.}}]{Gabrielli:2013hma}%
  \BibitemOpen
  \bibfield  {author} {\bibinfo {author} {\bibfnamefont {E.}~\bibnamefont
  {Gabrielli}}, \bibinfo {author} {\bibfnamefont {M.}~\bibnamefont
  {Heikinheimo}}, \bibinfo {author} {\bibfnamefont {K.}~\bibnamefont
  {Kannike}}, \bibinfo {author} {\bibfnamefont {A.}~\bibnamefont {Racioppi}},
  \bibinfo {author} {\bibfnamefont {M.}~\bibnamefont {Raidal}},  \emph
  {et~al.},\ }\href {\doibase 10.1103/PhysRevD.89.015017} {\bibfield  {journal}
  {\bibinfo  {journal} {Phys.Rev.}\ }\textbf {\bibinfo {volume} {D89}},\
  \bibinfo {pages} {015017} (\bibinfo {year} {2014})},\ \Eprint
  {http://arxiv.org/abs/1309.6632} {arXiv:1309.6632 [hep-ph]} \BibitemShut
  {NoStop}%
\bibitem [{\citenamefont {Farzinnia}\ \emph {et~al.}(2013)\citenamefont
  {Farzinnia}, \citenamefont {He},\ and\ \citenamefont {Ren}}]{1308.0295}%
  \BibitemOpen
  \bibfield  {author} {\bibinfo {author} {\bibfnamefont {A.}~\bibnamefont
  {Farzinnia}}, \bibinfo {author} {\bibfnamefont {H.-J.}\ \bibnamefont {He}}, \
  and\ \bibinfo {author} {\bibfnamefont {J.}~\bibnamefont {Ren}},\ }\href
  {\doibase 10.1016/j.physletb.2013.09.060} {\bibfield  {journal} {\bibinfo
  {journal} {Phys.Lett.}\ }\textbf {\bibinfo {volume} {B727}},\ \bibinfo
  {pages} {141} (\bibinfo {year} {2013})},\ \Eprint
  {http://arxiv.org/abs/1308.0295} {arXiv:1308.0295 [hep-ph]} \BibitemShut
  {NoStop}%
\bibitem [{\citenamefont {Carone}\ and\ \citenamefont
  {Ramos}(2013)}]{1307.8428}%
  \BibitemOpen
  \bibfield  {author} {\bibinfo {author} {\bibfnamefont {C.~D.}\ \bibnamefont
  {Carone}}\ and\ \bibinfo {author} {\bibfnamefont {R.}~\bibnamefont {Ramos}},\
  }\href {\doibase 10.1103/PhysRevD.88.055020} {\bibfield  {journal} {\bibinfo
  {journal} {Phys.Rev.}\ }\textbf {\bibinfo {volume} {D88}},\ \bibinfo {pages}
  {055020} (\bibinfo {year} {2013})},\ \Eprint {http://arxiv.org/abs/1307.8428}
  {arXiv:1307.8428 [hep-ph]} \BibitemShut {NoStop}%
\bibitem [{\citenamefont {Gorbunov}\ and\ \citenamefont
  {Tokareva}(2014)}]{1307.5298}%
  \BibitemOpen
  \bibfield  {author} {\bibinfo {author} {\bibfnamefont {D.}~\bibnamefont
  {Gorbunov}}\ and\ \bibinfo {author} {\bibfnamefont {A.}~\bibnamefont
  {Tokareva}},\ }\href {\doibase 10.1016/j.physletb.2014.10.036} {\bibfield
  {journal} {\bibinfo  {journal} {Phys.Lett.}\ }\textbf {\bibinfo {volume}
  {B739}},\ \bibinfo {pages} {50} (\bibinfo {year} {2014})},\ \Eprint
  {http://arxiv.org/abs/1307.5298} {arXiv:1307.5298 [astro-ph.CO]} \BibitemShut
  {NoStop}%
\bibitem [{\citenamefont {Khoze}\ and\ \citenamefont {Ro}(2013)}]{1307.3764}%
  \BibitemOpen
  \bibfield  {author} {\bibinfo {author} {\bibfnamefont {V.~V.}\ \bibnamefont
  {Khoze}}\ and\ \bibinfo {author} {\bibfnamefont {G.}~\bibnamefont {Ro}},\
  }\href {\doibase 10.1007/JHEP10(2013)075} {\bibfield  {journal} {\bibinfo
  {journal} {JHEP}\ }\textbf {\bibinfo {volume} {1310}},\ \bibinfo {pages}
  {075} (\bibinfo {year} {2013})},\ \Eprint {http://arxiv.org/abs/1307.3764}
  {arXiv:1307.3764} \BibitemShut {NoStop}%
\bibitem [{\citenamefont {Oda}(2013)}]{1305.0884}%
  \BibitemOpen
  \bibfield  {author} {\bibinfo {author} {\bibfnamefont {I.}~\bibnamefont
  {Oda}},\ }\href {\doibase 10.1016/j.physletb.2013.06.014} {\bibfield
  {journal} {\bibinfo  {journal} {Phys.Lett.}\ }\textbf {\bibinfo {volume}
  {B724}},\ \bibinfo {pages} {160} (\bibinfo {year} {2013})},\ \Eprint
  {http://arxiv.org/abs/1305.0884} {arXiv:1305.0884 [hep-ph]} \BibitemShut
  {NoStop}%
\bibitem [{\citenamefont {Orikasa}(2013)}]{1304.4700}%
  \BibitemOpen
  \bibfield  {author} {\bibinfo {author} {\bibfnamefont {Y.}~\bibnamefont
  {Orikasa}},\ }\href@noop {} {\bibfield  {journal} {\bibinfo  {journal} {CNUM:
  C13-02-13.1}\ } (\bibinfo {year} {2013})},\ \Eprint
  {http://arxiv.org/abs/1304.4700} {arXiv:1304.4700 [hep-ph]} \BibitemShut
  {NoStop}%
\bibitem [{\citenamefont {Fabbrichesi}\ and\ \citenamefont
  {Petcov}(2014)}]{1304.4001}%
  \BibitemOpen
  \bibfield  {author} {\bibinfo {author} {\bibfnamefont {M.}~\bibnamefont
  {Fabbrichesi}}\ and\ \bibinfo {author} {\bibfnamefont {S.}~\bibnamefont
  {Petcov}},\ }\href {\doibase 10.1140/epjc/s10052-014-2774-x} {\bibfield
  {journal} {\bibinfo  {journal} {Eur.Phys.J.}\ }\textbf {\bibinfo {volume}
  {C74}},\ \bibinfo {pages} {2774} (\bibinfo {year} {2014})},\ \Eprint
  {http://arxiv.org/abs/1304.4001} {arXiv:1304.4001 [hep-ph]} \BibitemShut
  {NoStop}%
\bibitem [{\citenamefont {Iso}\ and\ \citenamefont
  {Orikasa}(2013)}]{1210.2848}%
  \BibitemOpen
  \bibfield  {author} {\bibinfo {author} {\bibfnamefont {S.}~\bibnamefont
  {Iso}}\ and\ \bibinfo {author} {\bibfnamefont {Y.}~\bibnamefont {Orikasa}},\
  }\href {\doibase 10.1093/ptep/pts099} {\bibfield  {journal} {\bibinfo
  {journal} {PTEP}\ }\textbf {\bibinfo {volume} {2013}},\ \bibinfo {pages}
  {023B08} (\bibinfo {year} {2013})},\ \Eprint {http://arxiv.org/abs/1210.2848}
  {arXiv:1210.2848 [hep-ph]} \BibitemShut {NoStop}%
\bibitem [{\citenamefont {Hambye}\ and\ \citenamefont
  {Tytgat}(2008)}]{0707.0633}%
  \BibitemOpen
  \bibfield  {author} {\bibinfo {author} {\bibfnamefont {T.}~\bibnamefont
  {Hambye}}\ and\ \bibinfo {author} {\bibfnamefont {M.~H.}\ \bibnamefont
  {Tytgat}},\ }\href {\doibase 10.1016/j.physletb.2007.11.069} {\bibfield
  {journal} {\bibinfo  {journal} {Phys.Lett.}\ }\textbf {\bibinfo {volume}
  {B659}},\ \bibinfo {pages} {651} (\bibinfo {year} {2008})},\ \Eprint
  {http://arxiv.org/abs/0707.0633} {arXiv:0707.0633 [hep-ph]} \BibitemShut
  {NoStop}%
\bibitem [{\citenamefont {Foot}\ \emph {et~al.}(2008)\citenamefont {Foot},
  \citenamefont {Kobakhidze}, \citenamefont {McDonald},\ and\ \citenamefont
  {Volkas}}]{0709.2750}%
  \BibitemOpen
  \bibfield  {author} {\bibinfo {author} {\bibfnamefont {R.}~\bibnamefont
  {Foot}}, \bibinfo {author} {\bibfnamefont {A.}~\bibnamefont {Kobakhidze}},
  \bibinfo {author} {\bibfnamefont {K.~L.}\ \bibnamefont {McDonald}}, \ and\
  \bibinfo {author} {\bibfnamefont {R.~R.}\ \bibnamefont {Volkas}},\ }\href
  {\doibase 10.1103/PhysRevD.77.035006} {\bibfield  {journal} {\bibinfo
  {journal} {Phys.Rev.}\ }\textbf {\bibinfo {volume} {D77}},\ \bibinfo {pages}
  {035006} (\bibinfo {year} {2008})},\ \Eprint {http://arxiv.org/abs/0709.2750}
  {arXiv:0709.2750 [hep-ph]} \BibitemShut {NoStop}%
\bibitem [{\citenamefont {Shaposhnikov}\ and\ \citenamefont
  {Zenhausern}(2009)}]{0809.3406}%
  \BibitemOpen
  \bibfield  {author} {\bibinfo {author} {\bibfnamefont {M.}~\bibnamefont
  {Shaposhnikov}}\ and\ \bibinfo {author} {\bibfnamefont {D.}~\bibnamefont
  {Zenhausern}},\ }\href {\doibase 10.1016/j.physletb.2008.11.041} {\bibfield
  {journal} {\bibinfo  {journal} {Phys.Lett.}\ }\textbf {\bibinfo {volume}
  {B671}},\ \bibinfo {pages} {162} (\bibinfo {year} {2009})},\ \Eprint
  {http://arxiv.org/abs/0809.3406} {arXiv:0809.3406 [hep-th]} \BibitemShut
  {NoStop}%
\bibitem [{\citenamefont {Jaynes}(2003)}]{Jaynes:2003}%
  \BibitemOpen
  \bibfield  {author} {\bibinfo {author} {\bibfnamefont {E.~T.}\ \bibnamefont
  {Jaynes}},\ }\href@noop {} {\emph {\bibinfo {title} {{Probability theory: the
  logic of science}}}},\ edited by\ \bibinfo {editor} {\bibfnamefont {G.~L.}\
  \bibnamefont {Bretthorst}}\ (\bibinfo  {publisher} {Cambridge University
  Press},\ \bibinfo {address} {Cambridge, UK, New York},\ \bibinfo {year}
  {2003})\BibitemShut {NoStop}%
\bibitem [{\citenamefont {Gregory}(2005)}]{Gregory}%
  \BibitemOpen
  \bibfield  {author} {\bibinfo {author} {\bibfnamefont {P.}~\bibnamefont
  {Gregory}},\ }\href@noop {} {\emph {\bibinfo {title} {{Bayesian Logical Data
  Analysis for the Physical Sciences}}}},\ \bibinfo {edition} {1st}\ ed.\
  (\bibinfo  {publisher} {Cambridge University Press},\ \bibinfo {year}
  {2005})\BibitemShut {NoStop}%
\bibitem [{\citenamefont {Popper}(1959)}]{Popper}%
  \BibitemOpen
  \bibfield  {author} {\bibinfo {author} {\bibfnamefont {K.~R.}\ \bibnamefont
  {Popper}},\ }\href@noop {} {\emph {\bibinfo {title} {{The Logic of Scientific
  Discovery}}}}\ (\bibinfo  {publisher} {Routledge},\ \bibinfo {year}
  {1959})\BibitemShut {NoStop}%
\bibitem [{\citenamefont {Vickers}(2014)}]{sep-induction-problem}%
  \BibitemOpen
  \bibfield  {author} {\bibinfo {author} {\bibfnamefont {J.}~\bibnamefont
  {Vickers}},\ }in\ \href@noop {} {\emph {\bibinfo {booktitle} {{The Stanford
  Encyclopedia of Philosophy}}}},\ \bibinfo {editor} {edited by\ \bibinfo
  {editor} {\bibfnamefont {E.~N.}\ \bibnamefont {Zalta}}}\ (\bibinfo
  {publisher} {Stanford University},\ \bibinfo {year} {2014})\ \bibinfo
  {edition} {fall 2014}\ ed.\BibitemShut {Stop}%
\bibitem [{\citenamefont {Earman}(1992)}]{Earman}%
  \BibitemOpen
  \bibfield  {author} {\bibinfo {author} {\bibfnamefont {J.}~\bibnamefont
  {Earman}},\ }\href@noop {} {\emph {\bibinfo {title} {{Bayes or bust?: a
  critical examination of Bayesian confirmation theory}}}}\ (\bibinfo
  {publisher} {MIT Press},\ \bibinfo {year} {1992})\BibitemShut {NoStop}%
\bibitem [{\citenamefont {Lakatos}(1974)}]{Lakatos}%
  \BibitemOpen
  \bibfield  {author} {\bibinfo {author} {\bibfnamefont {I.}~\bibnamefont
  {Lakatos}},\ }\href@noop {} {\bibfield  {journal} {\bibinfo  {journal}
  {Studies in History and Philosophy of Science Part A}\ }\textbf {\bibinfo
  {volume} {4}},\ \bibinfo {pages} {309} (\bibinfo {year} {1974})}\BibitemShut
  {NoStop}%
\bibitem [{\citenamefont {Fowlie}(2014{\natexlab{a}})}]{Fowlie:2014xha}%
  \BibitemOpen
  \bibfield  {author} {\bibinfo {author} {\bibfnamefont {A.}~\bibnamefont
  {Fowlie}},\ }\href {\doibase 10.1103/PhysRevD.90.015010} {\bibfield
  {journal} {\bibinfo  {journal} {Phys.Rev.}\ }\textbf {\bibinfo {volume}
  {D90}},\ \bibinfo {pages} {015010} (\bibinfo {year} {2014}{\natexlab{a}})},\
  \Eprint {http://arxiv.org/abs/1403.3407} {arXiv:1403.3407 [hep-ph]}
  \BibitemShut {NoStop}%
\bibitem [{\citenamefont {Spade}\ and\ \citenamefont
  {Panaccio}(2011)}]{sep-ockham}%
  \BibitemOpen
  \bibfield  {author} {\bibinfo {author} {\bibfnamefont {P.~V.}\ \bibnamefont
  {Spade}}\ and\ \bibinfo {author} {\bibfnamefont {C.}~\bibnamefont
  {Panaccio}},\ }in\ \href@noop {} {\emph {\bibinfo {booktitle} {The Stanford
  Encyclopedia of Philosophy}}},\ \bibinfo {editor} {edited by\ \bibinfo
  {editor} {\bibfnamefont {E.~N.}\ \bibnamefont {Zalta}}}\ (\bibinfo {year}
  {2011})\ \bibinfo {edition} {fall 2011}\ ed.\BibitemShut {Stop}%
\bibitem [{\citenamefont {Jeffereys}\ and\ \citenamefont
  {Burger}(1991)}]{Burger}%
  \BibitemOpen
  \bibfield  {author} {\bibinfo {author} {\bibfnamefont {W.~H.}\ \bibnamefont
  {Jeffereys}}\ and\ \bibinfo {author} {\bibfnamefont {J.~O.}\ \bibnamefont
  {Burger}},\ }\href@noop {} {\emph {\bibinfo {title} {Sharpening Occam's Razor
  on a Bayesian Strop}}},\ \bibinfo {type} {Tech. Rep.}\ \bibinfo {number}
  {91-44C}\ (\bibinfo  {institution} {Department of Statistics, Purdue
  University},\ \bibinfo {year} {1991})\BibitemShut {NoStop}%
\bibitem [{\citenamefont {Ellis}\ \emph {et~al.}(1986)\citenamefont {Ellis},
  \citenamefont {Enqvist}, \citenamefont {Nanopoulos},\ and\ \citenamefont
  {Zwirner}}]{Ellis:1986yg}%
  \BibitemOpen
  \bibfield  {author} {\bibinfo {author} {\bibfnamefont {J.~R.}\ \bibnamefont
  {Ellis}}, \bibinfo {author} {\bibfnamefont {K.}~\bibnamefont {Enqvist}},
  \bibinfo {author} {\bibfnamefont {D.~V.}\ \bibnamefont {Nanopoulos}}, \ and\
  \bibinfo {author} {\bibfnamefont {F.}~\bibnamefont {Zwirner}},\ }\href
  {\doibase 10.1142/S0217732386000105} {\bibfield  {journal} {\bibinfo
  {journal} {Mod.Phys.Lett.}\ }\textbf {\bibinfo {volume} {A1}},\ \bibinfo
  {pages} {57} (\bibinfo {year} {1986})}\BibitemShut {NoStop}%
\bibitem [{\citenamefont {Barbieri}\ and\ \citenamefont
  {Giudice}(1988)}]{Barbieri:1987fn}%
  \BibitemOpen
  \bibfield  {author} {\bibinfo {author} {\bibfnamefont {R.}~\bibnamefont
  {Barbieri}}\ and\ \bibinfo {author} {\bibfnamefont {G.}~\bibnamefont
  {Giudice}},\ }\href {\doibase 10.1016/0550-3213(88)90171-X} {\bibfield
  {journal} {\bibinfo  {journal} {Nucl.Phys.}\ }\textbf {\bibinfo {volume}
  {B306}},\ \bibinfo {pages} {63} (\bibinfo {year} {1988})}\BibitemShut
  {NoStop}%
\bibitem [{\citenamefont {Allanach}\ \emph {et~al.}(2007)\citenamefont
  {Allanach}, \citenamefont {Cranmer}, \citenamefont {Lester},\ and\
  \citenamefont {Weber}}]{Allanach:2007qk}%
  \BibitemOpen
  \bibfield  {author} {\bibinfo {author} {\bibfnamefont {B.~C.}\ \bibnamefont
  {Allanach}}, \bibinfo {author} {\bibfnamefont {K.}~\bibnamefont {Cranmer}},
  \bibinfo {author} {\bibfnamefont {C.~G.}\ \bibnamefont {Lester}}, \ and\
  \bibinfo {author} {\bibfnamefont {A.~M.}\ \bibnamefont {Weber}},\ }\href
  {\doibase 10.1088/1126-6708/2007/08/023} {\bibfield  {journal} {\bibinfo
  {journal} {JHEP}\ }\textbf {\bibinfo {volume} {0708}},\ \bibinfo {pages}
  {023} (\bibinfo {year} {2007})},\ \Eprint {http://arxiv.org/abs/0705.0487}
  {arXiv:0705.0487 [hep-ph]} \BibitemShut {NoStop}%
\bibitem [{\citenamefont {Cabrera}\ \emph {et~al.}(2009)\citenamefont
  {Cabrera}, \citenamefont {Casas},\ and\ \citenamefont {Ruiz~de
  Austri}}]{Cabrera:2008tj}%
  \BibitemOpen
  \bibfield  {author} {\bibinfo {author} {\bibfnamefont {M.~E.}\ \bibnamefont
  {Cabrera}}, \bibinfo {author} {\bibfnamefont {J.~A.}\ \bibnamefont {Casas}},
  \ and\ \bibinfo {author} {\bibfnamefont {R.}~\bibnamefont {Ruiz~de Austri}},\
  }\href {\doibase 10.1088/1126-6708/2009/03/075} {\bibfield  {journal}
  {\bibinfo  {journal} {JHEP}\ }\textbf {\bibinfo {volume} {0903}},\ \bibinfo
  {pages} {075} (\bibinfo {year} {2009})},\ \Eprint
  {http://arxiv.org/abs/0812.0536} {arXiv:0812.0536 [hep-ph]} \BibitemShut
  {NoStop}%
\bibitem [{\citenamefont {Cabrera}\ \emph {et~al.}(2010)\citenamefont
  {Cabrera}, \citenamefont {Casas},\ and\ \citenamefont {Ruiz~de
  Austri}}]{Cabrera:2009dm}%
  \BibitemOpen
  \bibfield  {author} {\bibinfo {author} {\bibfnamefont {M.~E.}\ \bibnamefont
  {Cabrera}}, \bibinfo {author} {\bibfnamefont {J.~A.}\ \bibnamefont {Casas}},
  \ and\ \bibinfo {author} {\bibfnamefont {R.}~\bibnamefont {Ruiz~de Austri}},\
  }\href {\doibase 10.1007/JHEP05(2010)043} {\bibfield  {journal} {\bibinfo
  {journal} {JHEP}\ }\textbf {\bibinfo {volume} {1005}},\ \bibinfo {pages}
  {043} (\bibinfo {year} {2010})},\ \Eprint {http://arxiv.org/abs/0911.4686}
  {arXiv:0911.4686 [hep-ph]} \BibitemShut {NoStop}%
\bibitem [{\citenamefont {Fichet}(2012)}]{Fichet:2012sn}%
  \BibitemOpen
  \bibfield  {author} {\bibinfo {author} {\bibfnamefont {S.}~\bibnamefont
  {Fichet}},\ }\href {\doibase 10.1103/PhysRevD.86.125029} {\bibfield
  {journal} {\bibinfo  {journal} {Phys.Rev.}\ }\textbf {\bibinfo {volume}
  {D86}},\ \bibinfo {pages} {125029} (\bibinfo {year} {2012})},\ \Eprint
  {http://arxiv.org/abs/1204.4940} {arXiv:1204.4940 [hep-ph]} \BibitemShut
  {NoStop}%
\bibitem [{\citenamefont {Fowlie}(2014{\natexlab{b}})}]{Fowlie:2014faa}%
  \BibitemOpen
  \bibfield  {author} {\bibinfo {author} {\bibfnamefont {A.}~\bibnamefont
  {Fowlie}},\ }\href {\doibase 10.1140/epjc/s10052-014-3105-y} {\bibfield
  {journal} {\bibinfo  {journal} {Eur.Phys.J.}\ }\textbf {\bibinfo {volume}
  {C74}},\ \bibinfo {pages} {3105} (\bibinfo {year} {2014}{\natexlab{b}})},\
  \Eprint {http://arxiv.org/abs/1407.7534} {arXiv:1407.7534 [hep-ph]}
  \BibitemShut {NoStop}%
\bibitem [{\citenamefont {Baer}\ \emph {et~al.}(2013)\citenamefont {Baer},
  \citenamefont {Barger},\ and\ \citenamefont {Mickelson}}]{Baer:2013gva}%
  \BibitemOpen
  \bibfield  {author} {\bibinfo {author} {\bibfnamefont {H.}~\bibnamefont
  {Baer}}, \bibinfo {author} {\bibfnamefont {V.}~\bibnamefont {Barger}}, \ and\
  \bibinfo {author} {\bibfnamefont {D.}~\bibnamefont {Mickelson}},\ }\href
  {\doibase 10.1103/PhysRevD.88.095013} {\bibfield  {journal} {\bibinfo
  {journal} {Phys.Rev.}\ }\textbf {\bibinfo {volume} {D88}},\ \bibinfo {pages}
  {095013} (\bibinfo {year} {2013})},\ \Eprint {http://arxiv.org/abs/1309.2984}
  {arXiv:1309.2984 [hep-ph]} \BibitemShut {NoStop}%
\bibitem [{\citenamefont {Ghilencea}\ \emph {et~al.}(2012)\citenamefont
  {Ghilencea}, \citenamefont {Lee},\ and\ \citenamefont
  {Park}}]{Ghilencea:2012gz}%
  \BibitemOpen
  \bibfield  {author} {\bibinfo {author} {\bibfnamefont {D.~M.}\ \bibnamefont
  {Ghilencea}}, \bibinfo {author} {\bibfnamefont {H.~M.}\ \bibnamefont {Lee}},
  \ and\ \bibinfo {author} {\bibfnamefont {M.}~\bibnamefont {Park}},\ }\href
  {\doibase 10.1007/JHEP07(2012)046} {\bibfield  {journal} {\bibinfo  {journal}
  {JHEP}\ }\textbf {\bibinfo {volume} {1207}},\ \bibinfo {pages} {046}
  (\bibinfo {year} {2012})},\ \Eprint {http://arxiv.org/abs/1203.0569}
  {arXiv:1203.0569 [hep-ph]} \BibitemShut {NoStop}%
\bibitem [{\citenamefont {Ghilencea}\ and\ \citenamefont
  {Ross}(2013)}]{Ghilencea:2012qk}%
  \BibitemOpen
  \bibfield  {author} {\bibinfo {author} {\bibfnamefont {D.}~\bibnamefont
  {Ghilencea}}\ and\ \bibinfo {author} {\bibfnamefont {G.}~\bibnamefont
  {Ross}},\ }\href {\doibase 10.1016/j.nuclphysb.2012.11.007} {\bibfield
  {journal} {\bibinfo  {journal} {Nucl.Phys.}\ }\textbf {\bibinfo {volume}
  {B868}},\ \bibinfo {pages} {65} (\bibinfo {year} {2013})},\ \Eprint
  {http://arxiv.org/abs/1208.0837} {arXiv:1208.0837 [hep-ph]} \BibitemShut
  {NoStop}%
\bibitem [{\citenamefont {Ghilencea}(2014)}]{Ghilencea:2013nxa}%
  \BibitemOpen
  \bibfield  {author} {\bibinfo {author} {\bibfnamefont {D.}~\bibnamefont
  {Ghilencea}},\ }\href {\doibase 10.1103/PhysRevD.89.095007} {\bibfield
  {journal} {\bibinfo  {journal} {Phys.Rev.}\ }\textbf {\bibinfo {volume}
  {D89}},\ \bibinfo {pages} {095007} (\bibinfo {year} {2014})},\ \Eprint
  {http://arxiv.org/abs/1311.6144} {arXiv:1311.6144 [hep-ph]} \BibitemShut
  {NoStop}%
\bibitem [{\citenamefont {Ghilencea}(2013{\natexlab{a}})}]{Ghilencea:2013hpa}%
  \BibitemOpen
  \bibfield  {author} {\bibinfo {author} {\bibfnamefont {D.}~\bibnamefont
  {Ghilencea}},\ }\href {\doibase 10.1016/j.nuclphysb.2013.07.024} {\bibfield
  {journal} {\bibinfo  {journal} {Nucl.Phys.}\ }\textbf {\bibinfo {volume}
  {B876}},\ \bibinfo {pages} {16} (\bibinfo {year} {2013}{\natexlab{a}})},\
  \Eprint {http://arxiv.org/abs/1302.5262} {arXiv:1302.5262 [hep-ph]}
  \BibitemShut {NoStop}%
\bibitem [{\citenamefont {Antoniadis}\ \emph {et~al.}(2014)\citenamefont
  {Antoniadis}, \citenamefont {Babalic},\ and\ \citenamefont
  {Ghilencea}}]{Antoniadis:2014eta}%
  \BibitemOpen
  \bibfield  {author} {\bibinfo {author} {\bibfnamefont {I.}~\bibnamefont
  {Antoniadis}}, \bibinfo {author} {\bibfnamefont {E.}~\bibnamefont {Babalic}},
  \ and\ \bibinfo {author} {\bibfnamefont {D.}~\bibnamefont {Ghilencea}},\
  }\href {\doibase 10.1140/epjc/s10052-014-3050-9} {\bibfield  {journal}
  {\bibinfo  {journal} {Eur.Phys.J.}\ }\textbf {\bibinfo {volume} {C74}},\
  \bibinfo {pages} {3050} (\bibinfo {year} {2014})},\ \Eprint
  {http://arxiv.org/abs/1405.4314} {arXiv:1405.4314 [hep-ph]} \BibitemShut
  {NoStop}%
\bibitem [{\citenamefont {Ghilencea}(2013{\natexlab{b}})}]{Ghilencea:2013fka}%
  \BibitemOpen
  \bibfield  {author} {\bibinfo {author} {\bibfnamefont {D.}~\bibnamefont
  {Ghilencea}},\ }\href@noop {} {\bibfield  {journal} {\bibinfo  {journal}
  {PoS}\ }\textbf {\bibinfo {volume} {Corfu2012}},\ \bibinfo {pages} {034}
  (\bibinfo {year} {2013}{\natexlab{b}})},\ \Eprint
  {http://arxiv.org/abs/1304.1193} {arXiv:1304.1193 [hep-ph]} \BibitemShut
  {NoStop}%
\bibitem [{\citenamefont {Weinberg}(1989)}]{Weinberg:1988cp}%
  \BibitemOpen
  \bibfield  {author} {\bibinfo {author} {\bibfnamefont {S.}~\bibnamefont
  {Weinberg}},\ }\href {\doibase 10.1103/RevModPhys.61.1} {\bibfield  {journal}
  {\bibinfo  {journal} {Rev.Mod.Phys.}\ }\textbf {\bibinfo {volume} {61}},\
  \bibinfo {pages} {1} (\bibinfo {year} {1989})}\BibitemShut {NoStop}%
\bibitem [{\citenamefont {Wilson}\ and\ \citenamefont
  {Kogut}(1974)}]{Wilson:1973jj}%
  \BibitemOpen
  \bibfield  {author} {\bibinfo {author} {\bibfnamefont {K.}~\bibnamefont
  {Wilson}}\ and\ \bibinfo {author} {\bibfnamefont {J.~B.}\ \bibnamefont
  {Kogut}},\ }\href {\doibase 10.1016/0370-1573(74)90023-4} {\bibfield
  {journal} {\bibinfo  {journal} {Phys.Rept.}\ }\textbf {\bibinfo {volume}
  {12}},\ \bibinfo {pages} {75} (\bibinfo {year} {1974})}\BibitemShut {NoStop}%
\bibitem [{\citenamefont {Dirac}(1937)}]{Dirac:1937ti}%
  \BibitemOpen
  \bibfield  {author} {\bibinfo {author} {\bibfnamefont {P.~A.}\ \bibnamefont
  {Dirac}},\ }\href {\doibase 10.1038/139323a0} {\bibfield  {journal} {\bibinfo
   {journal} {Nature}\ }\textbf {\bibinfo {volume} {139}},\ \bibinfo {pages}
  {323} (\bibinfo {year} {1937})}\BibitemShut {NoStop}%
\bibitem [{\citenamefont {'t~Hooft}(1980)}]{'tHooft:1979bh}%
  \BibitemOpen
  \bibfield  {author} {\bibinfo {author} {\bibfnamefont {G.}~\bibnamefont
  {'t~Hooft}},\ }\href@noop {} {\bibfield  {journal} {\bibinfo  {journal} {NATO
  Sci.Ser.B}\ }\textbf {\bibinfo {volume} {59}},\ \bibinfo {pages} {135}
  (\bibinfo {year} {1980})}\BibitemShut {NoStop}%
\end{thebibliography}%
\end{document}